\documentclass[%
 reprint,
 table,
superscriptaddress,
 amsmath,amssymb,
 aps,
 pra,
]{revtex4-2}

\usepackage[utf8]{inputenc}
\usepackage[T1]{fontenc}
\usepackage{amssymb}
\usepackage{mathtools}
\usepackage{bbold}
\usepackage{bm} 
\usepackage[position=top, caption=false]{subfig}
\usepackage[table,x11names,dvipsnames,table]{xcolor}
\usepackage{tikz}
\usetikzlibrary {shapes.geometric}

\usepackage{amsthm}
\usepackage{amsmath}
\usepackage{mathtools}
\usepackage{mathrsfs}
\usepackage{physics}
\usepackage{nicefrac}
\usepackage{xfrac}
\usepackage{graphicx} 
\usepackage[pdftex, pdftitle={Article}, pdfauthor={Author}]{hyperref}
\usepackage[noabbrev]{cleveref}
\allowdisplaybreaks

\theoremstyle{definition}

\usepackage{dcolumn}

\begin{document}

\preprint{APS/123-QED}

\title{Entanglement teleportation along a regenerating hamster-wheel graph state}

\author{Haiyue Kang}
 \email{haiyuek@student.unimelb.edu.au}
 \affiliation{School of Physics, The University of Melbourne, Parkville, Victoria 3010, Australia}
 \affiliation{Centre for Quantum Biotechnology, School of Physics, The University of Melbourne\\
 Parkville, Victoria 3010, Australia}
 \author{John F. Kam}%
 \email{john.kam@monash.edu}
 \affiliation{School of Physics \& Astronomy, Monash University, Clayton, Victoria 3800, Australia}
\author{Gary J. Mooney}%
 \email{mooney.g@unimelb.edu.au}
\affiliation{School of Physics, The University of Melbourne, Parkville, Victoria 3010, Australia}
\author{Lloyd C. L. Hollenberg}%
 \email{lloydch@unimelb.edu.au}
\affiliation{School of Physics, The University of Melbourne, Parkville, Victoria 3010, Australia}
 \affiliation{Centre for Quantum Biotechnology, School of Physics, The University of Melbourne\\
 Parkville, Victoria 3010, Australia}

\date{\today}

\begin{abstract}
We scheme an efficient and reusable approach to quantum teleportation that allows cyclic teleportation of a two-qubit graph state around a quantum hamster wheel---a ring of qubits entangled as a one-dimensional line prepared on the 20-qubit Quantinuum H1-1 ion-trap quantum processor. The qubits on the ring are periodically measured and reused to achieve a teleportation depth that exceeds the total number of available qubits in the quantum processor. Using the outcomes measured during teleportation, we calculate and apply byproduct operators through dynamic circuits to correct local transformations induced on the teleported state. We evaluate the quality of teleportation by tracing the preserved entanglement and fidelity of the teleported two-qubit graph state from its density matrix. In the real-machine experiments, we demonstrate that 58\% of the teleported state's entanglement is sustained with a measured two-qubit negativity of $0.291\pm0.018$ after three complete revolutions around the hamster wheel, or equivalently, after hopping across 56 qubits. On the machine-specific noisy emulator, we found that the teleported state after 100 hops still sustained 45\% of its entanglement. By performing teleportation along a regenerating graph state, our work is a step forward in demonstrating the feasibility of measurement-based quantum computation. 

\end{abstract}

\maketitle


\begin{figure*}[t!]
    \centering
    \captionsetup{width=1\linewidth}
    \includegraphics[scale=0.72]{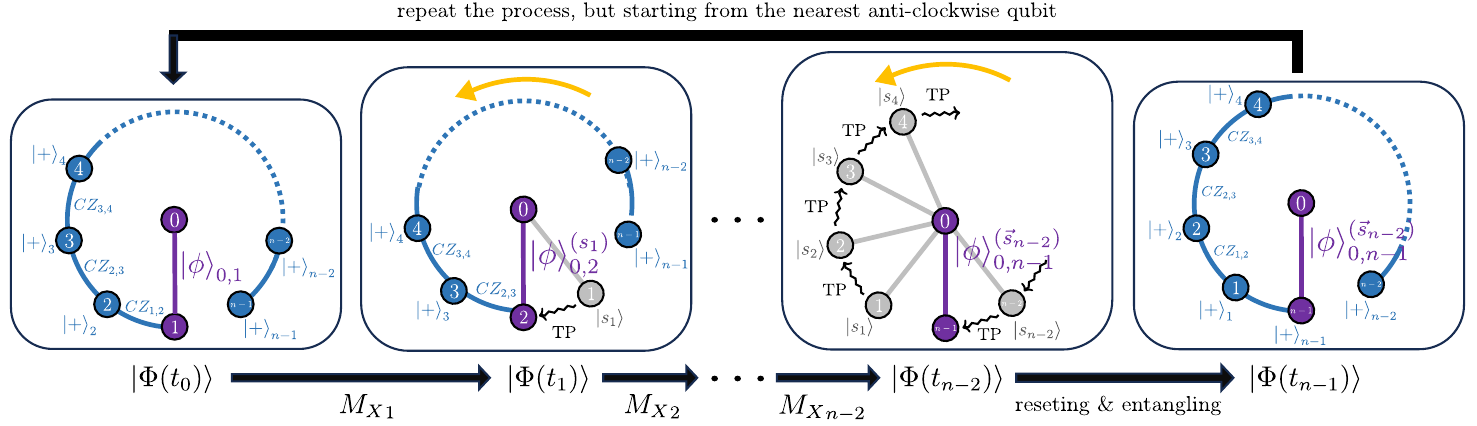}
    \caption{\label{fig:treadmill} The evolution of a general quantum hamster wheel consisting of $n$ qubits at different stages. In step 1, at the beginning $t_0$, the two-qubit state $\ket{\phi}_{0,1}$ is entangled with other qubits on the ring prepared in the Pauli-$X$ eigenstate $\ket{+}$ through $CZ$ gates. In step 2, by measuring qubit 1 to qubit $n-2$ in the Pauli-$X$ basis, the information on qubit 1 is teleported (TP) across the ring to qubit $n-2$, while maintaining its entanglement with qubit 0 up to some single-qubit transformation as a function of the measured results $\vec{s}_{n-2}$ of cardinality $n-2$, leaving an overall state $\ket{\Phi(t_{n-2})}=\ket{\vec{s}}\otimes(H^{n-2}Z^{s_1\oplus s_3\oplus \cdots}X^{s_2\oplus s_4\oplus\cdots})_{n-1}\ket{\phi}_{0,n-1}$. Orange arrows indicate the rotation direction of the `hamster wheel' from the perspective of the two-qubit state. In step 3, we reset all measured qubits and re-entangle the teleported two-qubit state with them in the same way as in $t_0$ but in the sequence starting from qubit 1 which is the first qubit measured in step 2. Repeating steps 2 and 3, the second qubit of state $\ket{\phi}$ is teleported around the ring of qubits with the number of hops more than available qubits.}
    \vspace{-10pt}
\end{figure*}
\emph{Introduction.}\label{sec:introduction}---As a natural inference derived from quantum entanglement, quantum teleportation is highly relevant to quantum information processing applications from computation to communication. Since the mechanism of quantum teleportation was first discovered two decades ago \cite{Teleportation}, there has been great interest in physical demonstrations of the phenomenon. In the early stages, researchers focused on empirical verification of basic protocols \cite{first_experimental_teleportation, teleportation_of_polarization_state}. As techniques advanced, researchers have successfully performed teleportation involving distant qubits \cite{teleportation_on_distant_qubits, teleportation_on_distant_qubits2, Haiyue_paper}. Apart from the physical phenomenon itself, potential applications that use the principle of quantum teleportation have also been investigated. These include dense coding \cite{dense_coding}
, quantum encrypted communication \cite{quantum_communication}, and detection of entanglement based on teleportation of the quantum state 
\cite{detect_entanglement_via_teleportation}. In the field of quantum computing, quantum teleportation can be exploited in hardware circuit compilation for constructing long-range Bell pairs, which can then be used to implement long-range multi-qubit gates 
\cite{long_range_gates_2, long_range_gates} based on the swapping of locally connected gates onto distant qubits through mid-circuit measurements, also known as entanglement swapping \cite{entanglement_swapping}. Quantum teleportation also plays a central role in the implementation of measurement-based quantum computation, which performs single-qubit measurements on entangled states that project the overall state equivalent to performing particular quantum gates \cite{dynamic_circuit, measurement_based_qc, measurement_based_qc2}. Despite these achievements, there are relatively few demonstrations of teleportation on entangled states across many qubits at a
level comparable to the current size of the state-of-the-art quantum devices. 

Here we demonstrate long-range quantum teleportation along a renewable resource of entangled qubits. For this we adapt the protocol outlined in \cite{Haiyue_paper}, to teleport a two-qubit entangled state (the ``hamster'') by 56 steps around a renewed graph state defined on a regenerating ring of 19 qubits (the ``wheel''). The protocol is based on repetitive
measurements on one qubit of two-qubit graph states such that the qubit is projected and propagating along the one-dimensional line of entangled qubits as in \cite{Haiyue_paper, dynamic_circuit}. However, since we are executing the teleportation along a walk that loops around the ring multiple times instead of a one-time path, we introduce additional layers of reset gates to reuse the teleported qubits before the next teleportation cycle. This solves the problem of the number of repeated nodes being more than the total number of qubits in a given quantum device. The evolution of the overall quantum state in our method thus uses principles similar to those used for measurement-based quantum computation \cite{measurement_based_qc2}. Therefore, our results could serve as an indicator for the potential utility of logical circuit depth in measurement-based quantum computation. The experiments were carried out
on the Quantinuum H1-1 quantum computer and the corresponding H1-1 emulator consisting of 20 ion-trapped qubits \cite{data_sheet}.

In the following contents, we cover the experimental setups and methodologies for implementing quantum hamster wheels, along with the mathematics of teleportation. Next, we introduce the concepts of negativity and fidelity as the two useful indicators of the amount of entanglement and the success of teleportation. Finally, we report and summarise the results obtained from the execution of quantum hamster wheels both on the emulator and on the real machine, which sustained the entanglement of the two-qubit state of high quality after 100 hops and 56 hops in teleportation, respectively.

\emph{Design of a Quantum Hamster Wheel.---\label{sec:methods}}The general procedure of setting up a quantum hamster wheel can be explained as a series of repeating entanglements and measurements. In the initial setup, we construct a state comprised of a fixed `axis' and a ring of nodes surrounding it. We prepare a pair of entangled qubits (qubit 0 and qubit 1) where qubit 0 always stays at the axis and qubit 1 will be at the top of the ring. Then, a one-dimensional line graph state is prepared on the ring of qubits that includes qubit 1, subsequent qubits to guide the teleportation path, and the final qubit at the end receiver site as shown in \Cref{fig:treadmill}. By measuring qubit 1 and its consecutive neighbours on the ring in the Pauli-$X$ basis, the second-qubit component of the two-qubit entangled pure state $\ket{\phi}_{0,1}\in \mathcal{H}_0\otimes\mathcal{H}_1$ teleports (TP) around the ring of qubits prepared in $\ket{+}$ states entangled via Controlled$-Z$ ($CZ$) gates while preserving its entanglement with qubit 0 on the axis. However, after every `hop' in teleportation, the new teleported two-qubit state will induce a transformation $U\in \mathcal{U}(\mathcal{H}_1)$ acted on the second qubit as a function of the measurement results from prior teleportations. Once the component of $\ket{\phi}_{0,1}$ on qubit 1 is teleported to the last qubit ($n-1$) along the ring, we perform resets on all of the measured qubits and re-entangle them with the last qubit back into a ring in the order when they were measured so that they can be reused for further teleportations as illustrated in \Cref{fig:treadmill}. By repeating such a process, the qubit 1 component of $\ket{\phi}_{0,1}$ can be teleported around the ring of qubits over multiple revolutions as a `quantum hamster wheel', extending the number of hops in teleportation beyond the total number of qubits given by a quantum device. In other words, the teleportation routine is equivalent to a \textit{walk} that constitutes multiple cycles. However, in more general cases, the all-to-all qubit connectivity of Quantinuum System Model H1 allows the sequence of qubits in those cycles to be arbitrary.

Here, we use the two-qubit graph state as the state $\ket{\phi}_{0,1}$ to be teleported around the hamster wheel. A graph state is constructed based on a graph $G=(V,E)$, where $V$ is the set of vertices and $E$ is the set of edges,
\begin{equation}
    \ket{\phi(G)}\coloneqq \prod\limits_{(u, v)\in E}{CZ_{(u, v)}{\ket{+}}^{\otimes \abs{V}}}.
\end{equation}
$\abs{V}$ is the number of qubits or vertices in the graph, and $CZ_{(u, v)}$ is the controlled-$Z$ gate acting on edge $e=(u, v)$ of the graph $G$. Thus, the two-qubit graph state $\ket{\phi(P_2)}$ defined on the path of two vertices $P_2$ initially prepared on qubit 0 and qubit 1 has the closed form
\begin{equation}
    \ket{\phi}_{0,1}\rightarrow\ket{\phi(P_{2})}_{0,1}\coloneqq\frac{1}{2}(\ket{00}+\ket{01}+\ket{10}-\ket{11})_{0,1}.
\end{equation}
The fact that it is being maximally entangled establishes a high starting point of the entanglement to be maintained when subject to noise from all sources during teleportation. Ultimately, a general $n$-qubit overall state $\ket{\Phi(t_0)}$ before teleportation comprises $\ket{\phi}_{0,1}$ and the chain of entangled qubits is a path graph state mapped from path $P_{n}$, $(n>2)$, where $V(P_{n})=\{0,1\cdots,n-1\}$, $E(P_{n})=\{(0,1),(1,2),\cdots(n-2,n-1)\}$. Hence, $\ket{\Phi(t_0)}$ is given by
\begin{align}
    \ket{\Phi(t_0)}&=\left(\prod\limits_{i=1}^{n-2}{CZ_{i,i+1}}\right)\bigotimes\limits_{j=2}^{n-1}\ket{+}_{j}\otimes\ket{\phi(P_2)}_{0,1}\nonumber\\
    &=\frac{1}{\sqrt{2^{n}}}\sum\limits_{\vec{x}\in \{0,1\}^{\otimes n}}{(-1)^{\sum\limits_{i=0}^{n-2}{x_i x_{i+1}}}\ket{\vec{x}}},
\end{align}
where $\ket{\vec{x}}=\ket{x_{0},x_{1},\cdots, x_{n-1}}$, and $x_{i}\in\{0,1\}$. The parity of the sum of $x_{i}x_{i+1}$ from $i=0$ to $i=n-2$ gives the phase of the component $\ket{\vec{x}}$ as a result of the $CZ$ gates applied to all nearest neighbour qubit pairs. In this paper, we set $n=20$ because it is the number of qubits available on the Quantinuum H1-1 computer unless otherwise specified.

\begin{figure*}[t!]
    \captionsetup{width=1\linewidth}
    \includegraphics[width=0.9\linewidth]{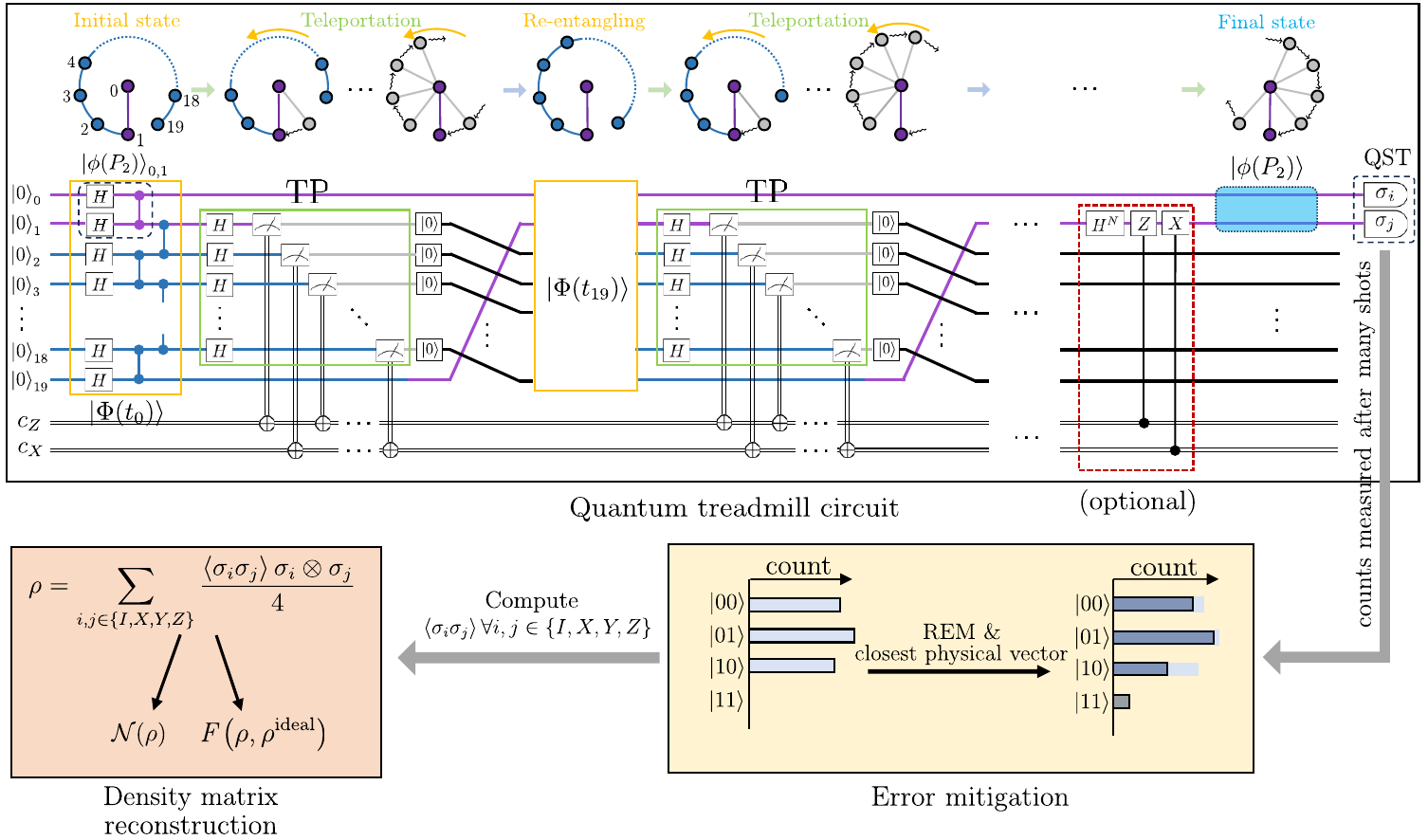}
    \caption{\label{fig:flowchart} Overall workflow of the quantum hamster wheel. Initially, the two-qubit graph state $\ket{\phi(P_2)}_{0,1}$ is entangled with other qubits as a line graph state from qubit 0 to qubit 19.  Next, the teleportation (TP) is performed by measuring the intermediate qubits in the Pauli-$X$ basis. The purple coloured wires indicate which qubits the two-qubit state is currently teleported to. The measured outcomes are binary added and stored in the classical registers $c_Z$ and $c_X$ which could (optionally) be used as the control for dynamic circuit correction later. After this, we reset all measured qubits, then re-entangle the two-qubit state with other qubits in the same way as the beginning. The crossing on the quantum wires only reshuffles the qubit displaying order so no SWAP gates are applied. We repeat the process a few times on the ongoing circuit, making the second qubit of $\ket{\phi(P_2)}$ teleport around the ring of qubits periodically. At the end of the circuit, the initial two-qubit state $\ket{\phi(P_2)}$ can be recovered by applying byproduct operators $(H^{m}Z^{s_{1}\oplus s_{3}\oplus\cdots}X^{s_{2}\oplus s_{4}\oplus\cdots})^{\dagger}$ ($s_i$ is the measured outcome of the $i$th hop and $m$ is the total number of hops) dynamically based on the mid-circuit measurement results. Alternatively, one can avoid dynamic circuits by using the mid-circuit measurement results to categorise the teleported state into one of the four possibilities via post-processing. The final two-qubit state is measured in the $3^2$ combinations of Pauli bases to perform Quantum State Tomography (QST). The bit-string counts measured from the QST are then passed to Readout Error Mitigation (REM) and the nearest physical probability vector algorithm to mitigate classical readout errors while being robust to imperfect REM calibration. Finally, we evaluate the entanglement and fidelity from the teleported two-qubit state's density matrix.}
    \vspace{-10pt}
\end{figure*}

After multiple revolutions of teleportation rotated around the hamster wheel with a total of $m$ hops, the teleported two-qubit state accumulates the net local transformation as a function of all measurement outcomes $\vec{s}$ that can be expressed by at most three gates,
\begin{equation}\label{eqn:byproduct_operators}
    \ket{\phi(P_{2})}^{(\vec{s},m)}_{0,r}\coloneqq I_{0}\otimes(H^{m}Z^{s_{1}\oplus s_{3}\oplus\cdots}X^{s_{2}\oplus s_{4}\oplus\cdots})_{r}\ket{\phi(P_{2})}_{0,r},
\end{equation}
where $\vec{s}=(s_1,s_2,\cdots,s_m)^T$ is the vector of all measured outcomes from the first hop to the last $m^{\text{th}}$ hop, $r$ is the label of the qubit that teleportation terminates calculated as $r=m\mod{19} + 1$, and $X,Z,H$ are the Pauli $X$, Pauli $Z$ and Hadamard gates respectively. Moreover, the symbol $\oplus$ represents binary addition, in other words, exclusive OR (XOR). From \Cref{eqn:byproduct_operators}, one could easily show that the teleported state $\ket{\phi(P_2)}_{0,r}^{(\vec{s},m)}$ only has four possible configurations of the teleported state that are locally-transformed from the original two-qubit state regardless of the measurement outcomes. Such a transformation is corrected by applying the byproduct operator $(H^{m}Z^{s_{1}\oplus s_{3}\oplus\cdots}X^{s_{2}\oplus s_{4}\oplus\cdots})^{\dagger}_{r}$ through dynamic circuits. The byproduct operator has its Pauli-$Z$ and Pauli-$X$ powers controlled by the discriminator that has been used in \cite{discriminant_vec, long_range_gates, Haiyue_paper},
\begin{equation}\label{eqn:discriminator}
\begin{aligned}
D(\vec{s})=\left(\bigoplus\limits_{i\in\text{odd}}{s_{i}},\bigoplus\limits_{j\in\text{even}}{s_{j}}\right),
\end{aligned}
\end{equation}
which evaluates the binary sums of measured outcomes corresponding to odd and even sequences of hops respectively. Apart from applying the dynamic circuits, an alternative is to bucket the teleported state into one of the four categories according to the discriminator by post-selection. Specific outcome states can be obtained by repeating until success. The general flow of the process, including error mitigation and density matrix reconstruction procedures, is illustrated in \Cref{fig:flowchart}.

\emph{Teleportation quality witnesses.---\label{sec:Entanglement Measurements}}As outlined in \Cref{fig:flowchart}, the quality of teleportation is characterised in two ways: negativity, which evaluates the amount of entanglement between two qubits of the teleported two-qubit state and fidelity, which accesses the proximity of the teleported state to its ideal counterpart.

\text{Negativity:}\hspace{1em}The negativity $\mathcal{N}$ between two partitioned quantum subsystems $a$ and $b$ is defined over their joint density matrix $\rho$ acting on the Hilbert space $\mathcal{H}_a\otimes\mathcal{H}_b$,
\begin{equation}\label{eq: negativity}
    \mathcal{N}(\rho)=\frac{1}{2}(||\rho^{T_{a}}||-1)=\abs{\sum\limits_{\lambda_{i}<0}{\lambda_{i}}},
\end{equation}
where $\rho^{T_{a}}$ denotes the partial transpose \cite{partial_transpose} of $\rho$ with respect to partition $a$, and $\lambda_{i}$ are the corresponding negative eigenvalues with respect to $\rho^{T_{a}}$. The negativity ranges from 0 to 0.5. In the special case for partitioning two qubits, the two partitions are maximally entangled when $\mathcal{N}=0.5$, and $\mathcal{N}>0$ is a necessary and sufficient condition to demonstrate the existence of entanglement \cite{414_fidel_paper, bipartite_65, 20q_entanglement}.


Fidelity:\hspace{1em}On the other hand, fidelity calculates the overlap between the noisy and ideal states. Specifically, when the noisy state is potentially mixed but the ideal state is pure, the fidelity between noisy and ideal states with density matrices $\rho$ and $\sigma$ respectively can be simplified as
\begin{equation}
    \textit{F}\left( \rho,\sigma \right)=\Bigg(\text{tr}\left(\sqrt{\sqrt{\sigma}\rho\sqrt{\sigma}}\right)\Bigg)^2\xrightarrow[\text{tr}(\sigma)=1]{}\text{tr}\left( \rho\sigma\right),
\end{equation}
where the properties $\sigma=\ket{\psi_{\sigma}}\bra{\psi_{\sigma}}$, $\sigma^2=\sigma$ and $\text{tr}(\sigma^2)=1$ for $\sigma$ as being a pure state was used in the derivation \cite{fidelity}.

\begin{figure*}[t!]
    \centering
    \subfloat[\label{fig:negativity QREM}Negativity]{
         \centering
         \:\includegraphics[width=0.85\columnwidth]{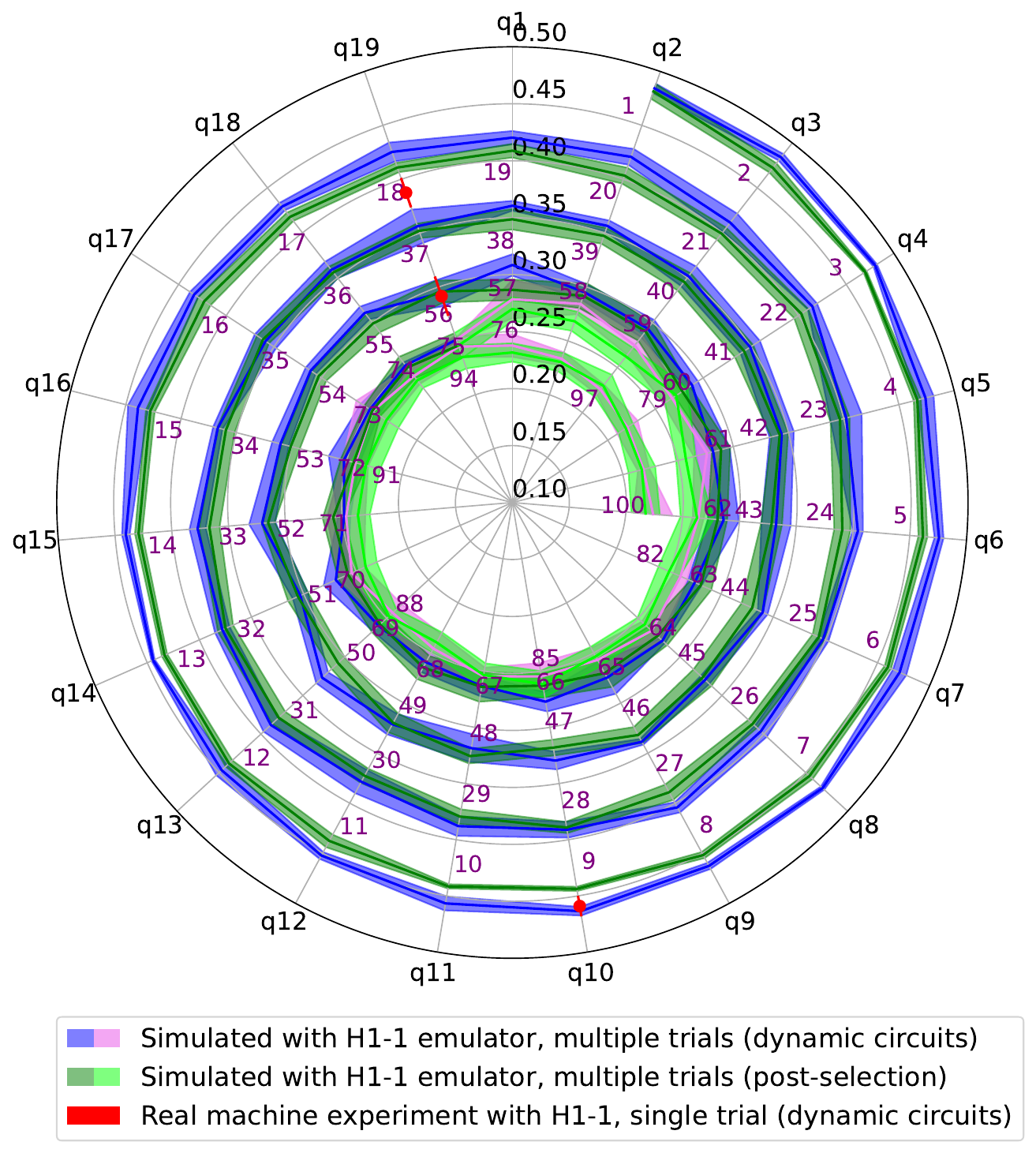}\:}
    \subfloat[\label{fig:fidelity QREM}Fidelity]{
         \centering
         \:\includegraphics[width=0.85\columnwidth]{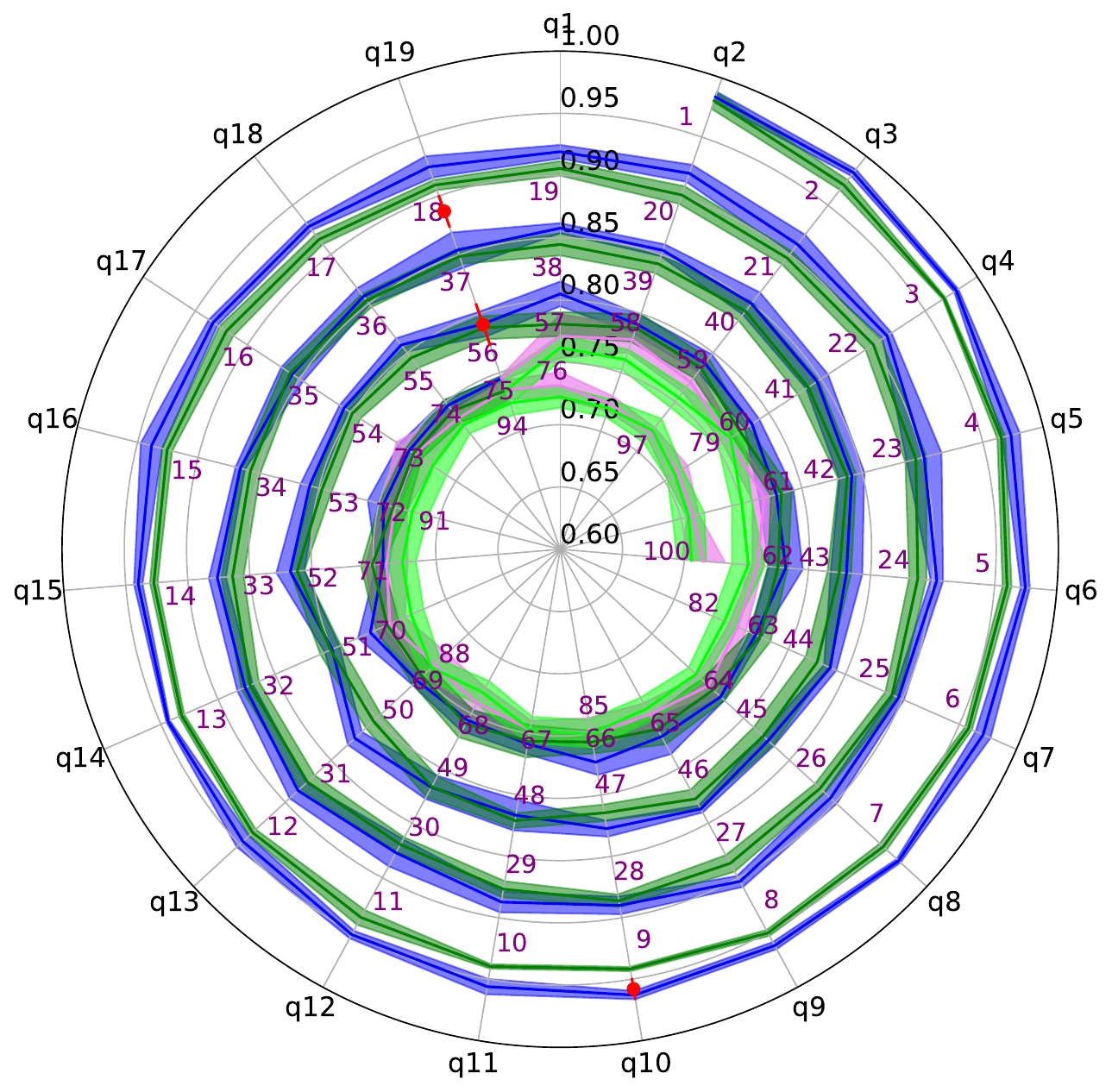}\:}
        \caption{Polar plots of the teleported two-qubit graph state showing \textbf{(a)} negativity and \textbf{(b)} fidelity on the radial axis versus the number of teleportation hops in the angular axis sampled from the H1-1 emulator averaged over four trials. The results for larger than 75 hops are marked using different colours to highlight the change in the measured entanglement due to different noise parameters arising from a time gap between experiments. The three data points marked in red represent results obtained from the physical H1-1 quantum computer for 9, 18 and 56 hops with respective negativities $0.459\pm0.009, 0.388\pm0.014, 0.291\pm0.018$ and fidelities $0.958\pm0.009, 0.887\pm0.014, 0.791\pm0.018$, where experiments were conducted on 21 March, 25 March and 24 April 2024 respectively. Assuming the distribution is Gaussian, the half-width of the shaded strip is two times the standard error of the sampled negativity or fidelity to represent 95\% confidence interval. On the other hand, the data obtained from real H1-1 quantum computers are only sampled in a single trial due to limited resources. Therefore, we use the technique of bootstrapping to resample the data from the measured sample distribution for 200 trials. The error bars represent 95\% confidence interval of the bootstrapped negativities or fidelities. All results are mitigated with REM as indicated in \textit{Teleportation quality witnesses}. Due to the feedforward gates applied based on mid-circuit measurements in the dynamic circuit approach, REM is applied to only the two qubits of the final teleported state. The corresponding plot with regular crossing axes are included in \Cref{fig:square plot}.}
    \label{fig:result}
\end{figure*}

Density matrix construction:\hspace{1em}To obtain these quantities as outlined above, we first employ Quantum State Tomography (QST) to reconstruct the density matrix of the two-qubit teleported state \cite{nielsen_chuang_2022} before evaluating entanglements. During this process, we use Readout Error Mitigation (REM) \cite{QREM} and the nearest physical probability vector algorithm proposed by Michelot \cite{closest_pvec} as described in \Cref{fig:flowchart} to calibrate the measured bit-string counts subject to readout bit-flip errors and eliminate any unphysical negative counts induced by imperfect calibration which could be caused by device drift and statistical errors. In future work, an alternative REM approach \cite{bo_yang_qrem} could be used for better results with more efficient time costs. To avoid the exponential scaling of calibration matrix dimension with respect to the number of measured qubits when applying the post-selection method, we apply REM qubit-wisely \cite{Mooney_2021} to restrict their dimensions to $2\times2$ for each iteration. Additionally, we use the algorithm \cite{closest_physical_rhos} by Smolin et al. to find the nearest physical density matrix which may originally contain unphysical negative eigenvalues.

Moreover, for the post-selection method, since it would randomly project the teleported state into one of the four variants at some given number of hops as indicated in \Cref{eqn:byproduct_operators}, we construct the density matrices for each variant separately in the form of
\begin{equation}
    \rho^{(\text{mixed})}=\frac{1}{2^{m-2}}\sum\limits_{\vec{s}}{\ket{\phi(P_{2})}^{(\vec{s},m)}\bra{\phi(P_{2})}^{(\vec{s},m)}}.
\end{equation}
Hence, the negativity and fidelity for each variant of the teleported state are individually evaluated, and their mean is taken to estimate the average quality of entanglement.

\emph{Results.---\label{sec:results}}Following the methods outlined in previous paragraphs, we characterise the negativity and fidelity of the teleported two-qubit state using a real quantum device (Quantinuum H1-1 quantum computer) together with a realistic device-noise model emulator (H1-1 emulator) together with the corresponding physical quantum computer (H1-1). The source code generated for this program are available in the GitHub repository \cite{Program_Codes}.

In \Cref{fig:negativity QREM} and \Cref{fig:fidelity QREM}, we present the negativities and fidelities recorded from the real machine (red-marked data points) and the emulator (blue data), plotted in polar plots against the number of hops that the two-qubit graph states were teleported across using the teleportation strategies. The emulated data is averaged over four trials. Due to limited access to computational resources, experiments on the H1-1 quantum computer were conducted for only three scenarios of 9, 18, and 56 hops in a single trial of the QST experiment with 1000 shots for each QST circuit. We bootstrap 200 samples from each QST basis measurement for both negativities and fidelities. The error bars of the data obtained from the H1-1 quantum computer of size $\epsilon$ take the half-range of the 95\% confidence interval of the bootstrapped negativities or fidelities, such that $\epsilon = \frac{1}{2} {\left(\left\lfloor a_{0.975N}\right\rfloor-\left\lceil a_{0.025N}\right\rceil\right)}$, where $\{a_n\},n=1,2,\cdots,N$ is the sequence of bootstrapped negativities or fidelities ordered from lowest to highest and $N=200$ is the sample size.
For both the negativity and fidelity, we have used REM, the nearest physical probability vector, and the nearest physical density matrix algorithm as previously outlined to mitigate classical readout errors. After all, from the H1-1 quantum computer we obtain the negativity of $0.459\pm0.009, 0.388\pm0.014, 0.291\pm0.018$ and fidelity of $0.958\pm0.009, 0.887\pm0.014, 0.791\pm0.018$ corresponds to 9, 18 and 56 hops respectively.  For the data obtained from the emulator, we report a clear trend that the entanglement of the teleported state decreases gradually, almost linearly, as the number of hops increases. This trend is closely in agreement with the data points obtained from the real device without significant deviations, as well as the trend of the two-qubit gate fidelity plot in the H1 model product data sheet \cite{data_sheet}. Therefore, it reinforces the reliability of the H1-1 emulator to approximate the corresponding device noise efficiently. There is a clear change that occurs at 76 hops, which results in a global increase in the measured negativity and fidelity from this point. Given the long time gap between experiments performed before and after the 75-hop mark, this may indicate the changes in noise model parameters used in the emulator. From the plot, we also find that both dynamic circuit and post-selection approaches exhibit very similar behaviour regarding the amount of entanglement. This observation agrees with our expectations that the two approaches should only introduce little differences in the negativity and fidelity, given that their circuit depth differs by at most three gates. On the other hand, we have shown a distinct utility of the post-selection approach against dynamic circuits presented in previous work \cite{Haiyue_paper} using IBM quantum computers. 
This was a direct consequence of the accumulated latency for processing each feed-forward mid-circuit measurement sequentially 
and the relatively high level of noise in IBM quantum computers. From these results, we confirm that even after 56 hops, the negativity of the state teleported using a dynamic circuit can be preserved up to $0.291\pm 0.018$. When tested using the emulator, it still maintains a value of $0.224\pm 0.009$ up to 100 hops, which is just over 5 full cycles. Therefore, it is reasonable to expect that the actual limit of the number of hops the two-qubit state can teleport across using H1-1 quantum computer while preserving entanglement is far beyond 100.

\emph{Discussion.---\label{sec:Discussions}}In summary, we demonstrated the teleportation of two-qubit graph states cyclically on a `quantum hamster wheel'. By making use of the reset gates, we were able to reuse the measured qubits and perform teleportation hops more than the total number of qubits available in the Quantinuum 20-qubit H1-1 quantum computer with ion-trap architecture. Despite limited real machine resources, in the emulation, we still showed that the entanglement of the two-qubit state was preserved with the negativity of $0.291\pm0.018$ after 56 hops in teleportation on the real quantum processor and $0.224\pm0.009$ after 100 hops on the corresponding emulator with the device-noise model. We expect this figure to be extended above 100 without the break-off of entanglement. Such performance benchmarks the utility of ion-trap quantum processors in achieving high-fidelity two-qubit gates and long coherence times that are more robust against perturbations.  Moreover, since the teleportation protocol with dynamic circuits shares a similar principle and process as the single-qubit measurement-based quantum computation, these results also serve as a feasibility demonstration of measurement-based quantum computation.
\medskip

\emph{Acknowledgements.---}This work was supported by the Australian Research Council Centre of Excellence for Quantum Biotechnology (CE230100021). The authors gratefully acknowledge the support of the University of Melbourne’s Zero Emission Energy Laboratory (ZEE Lab) and the Victorian Higher Education State Investment Fund (VHESIF). JFK was supported by an Australian Government Research Training Program Scholarship.

\emph{Data Availability Statement.---}
The source code and datasets generated and / or analysed during the current study are available from the corresponding author upon reasonable request.



\renewcommand{\href}[2]{#2}
\bibliography{reference}
\onecolumngrid
\appendix
\section*{End Matter}
We include two supplementary plots which describe the same information as in \Cref{fig:result} but plotted in regular axes layout.
\begin{figure}[h]
    \centering
    \subfloat[\label{fig:negativity QREM square}Negativity]{
         \centering
         \:\includegraphics[width=0.5\columnwidth]{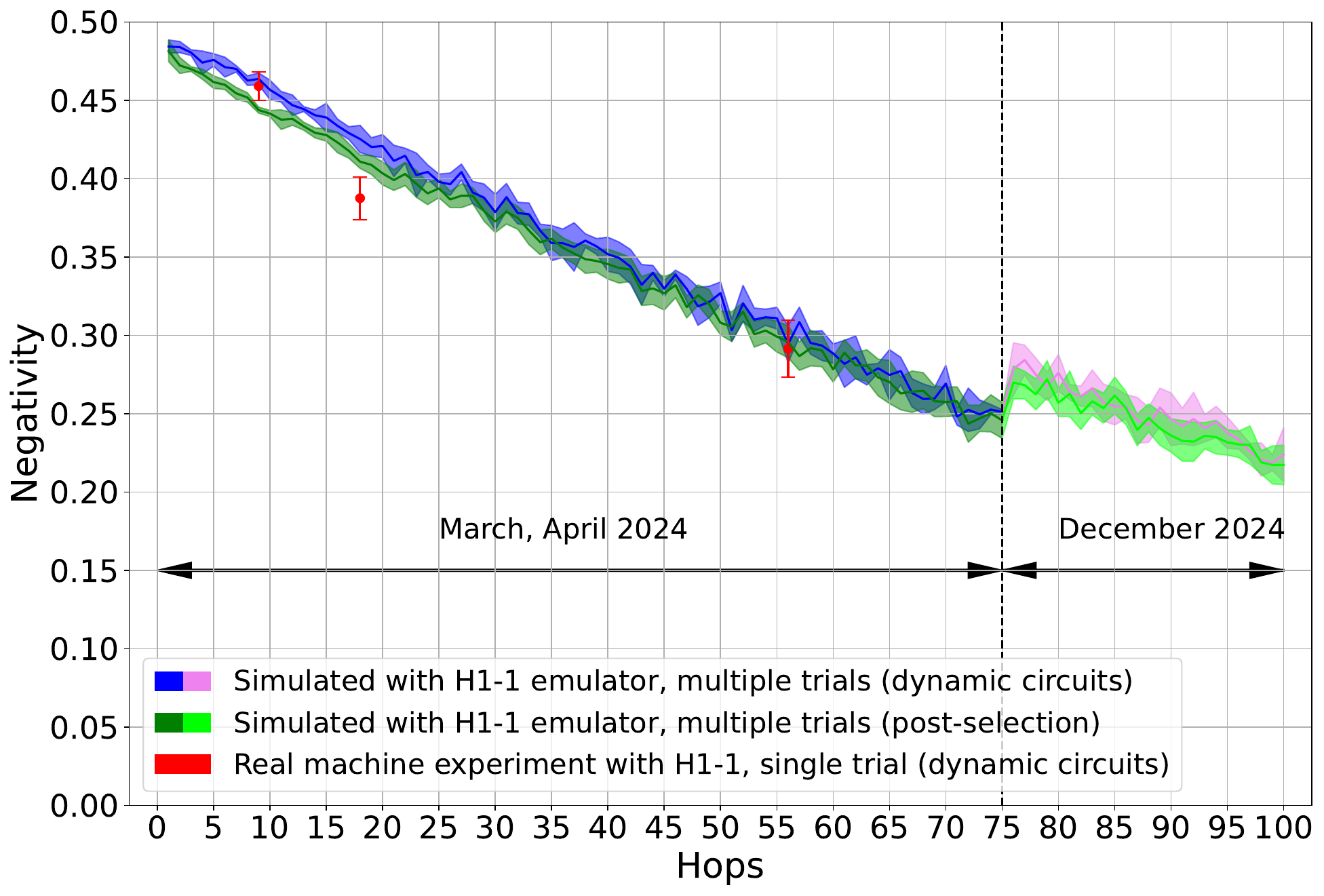}\:}
    \subfloat[\label{fig:fidelity QREM square}Fidelity]{
         \centering
         \:\includegraphics[width=0.5\columnwidth]{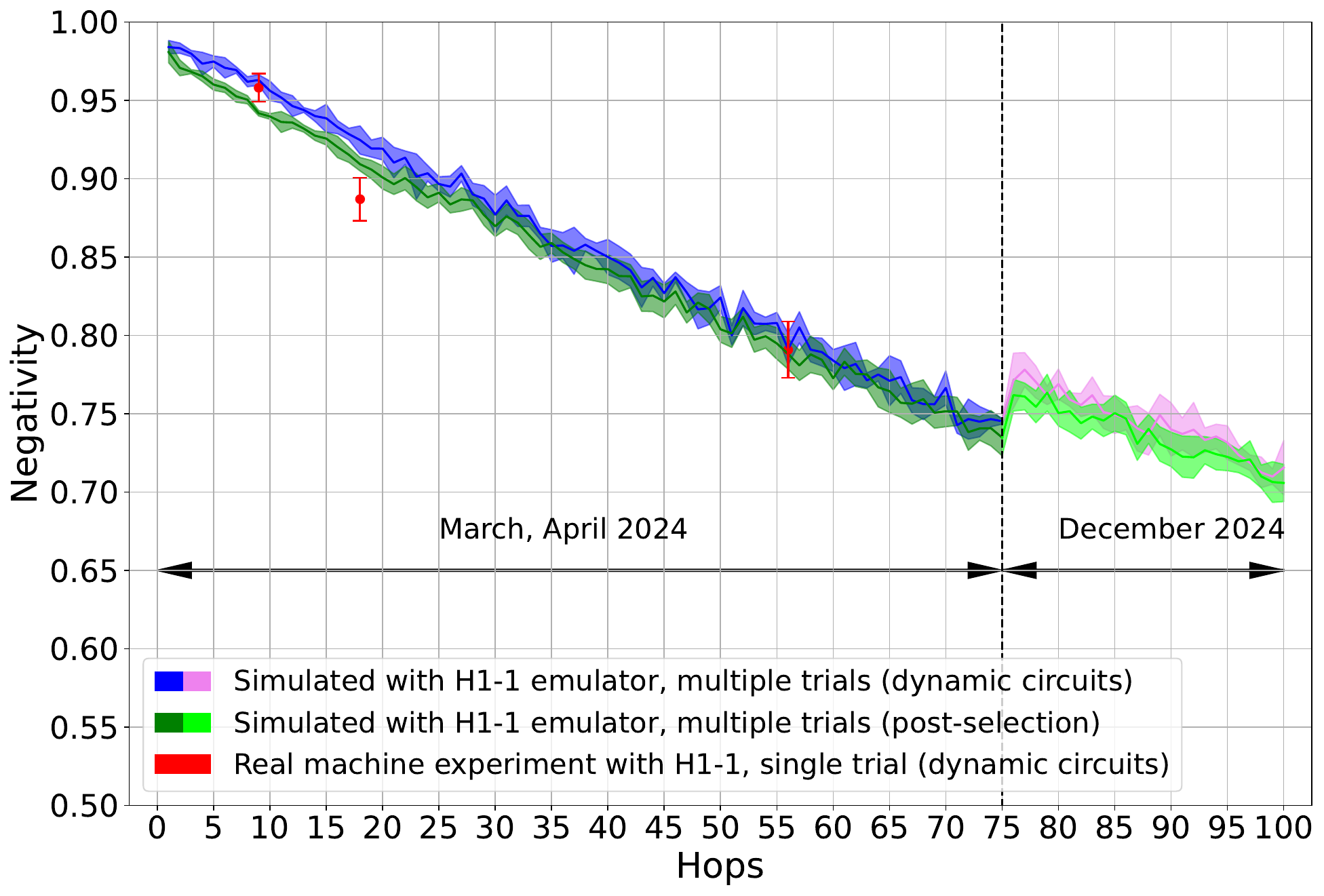}\:}
    \caption{The same \textbf{(a)} negativity and \textbf{(b)} fidelity plot as in \Cref{fig:result}, but plotted in the regular plot with perpendicular axes.}
    \label{fig:square plot}
\end{figure}
\end{document}